\documentclass[12pt]{iopart}

\usepackage{graphicx}

\newcommand{\Deff}{D_{\mathrm{eff}}}
\newcommand{\mz}{m_{\mathrm{z}}}
\newcommand{\tauzi}{\tau_{\mathrm{zi}}}
\newcommand{\nuzi}{\nu_{\mathrm{zi}}}

\begin{document}

\title[Collisional impurity transport and sheared electric fields]{Full orbit simulations of collisional impurity transport in spherical tokamak plasmas with strongly-sheared electric fields}
\author{C G Wrench$^1$, E Verwichte$^1$ and K G McClements$^2$}
\address{$^1$ Centre for Fusion, Space and Astrophysics, University of Warwick, Coventry, CV4 7AL, UK}
\address{$^2$ EURATOM/CCFE Fusion Association, Culham Science Centre, Abingdon, Oxfordshire, OX14 3DB, UK}
\ead{c.g.wrench@warwick.ac.uk}

\begin{abstract}
The collisional dynamics of test impurity ions in spherical tokamak plasmas with strongly-sheared radial electric fields is investigated by means of a test particle full orbit simulation code. The strength of the shear is such that the standard drift ordering can no longer be assumed and a full orbit approach is required. The effect of radial electric field shear on neoclassical particle transport is quantified for a range of test particle mass and charge numbers and electric field parameters. It is shown that the effect of a sheared electric field is to enhance the confinement of impurity species above the level observed in the absence of such a field. The effect may be explained in terms of a collisional drag force drift, which is proportional to particle charge number but independent of particle mass. This drift acts inwards for negative radial electric fields and outwards for positive fields, implying strongly enhanced confinement of highly ionized impurity ions in the presence of a negative radial electric field.
\end{abstract}

 %See: http://www.aip.org/pacs/pacs2010/individuals/pacs2010_regular_edition/reg50.htm#52

\submitto{Nuclear Fusion}

\maketitle

\section{Introduction}

Impurity ions present in tokamak plasmas can significantly degrade plasma performance through radiative losses and fuel dilution \cite{IPB99}. Conversely, impurity ions present in the edge or near the divertor of a tokamak may be beneficial by radiating thermal energy and thereby mitigating the heat load incident on the divertor and plasma facing components \cite{Murakami03}. An understanding of impurity transport (a topic which has received less attention than the transport of bulk ions) is crucial in order to predict the overall performance of fusion plasmas \cite{IPB99}.

The analysis of charged particle motion in strongly-magnetized plasmas, for example those in tokamaks, generally relies heavily on a multi-timescale approach \cite{Morozov66}, whereby the particle motion is decomposed into fast gyromotion about the magnetic field, streaming motion along the field (characterised by adiabatic invariance of the particle magnetic moment) and a much slower drift motion of the particle guiding centre across the field. This treatment is valid provided that the drift ordering is satisfied. Specifically, the particle Larmor radius must be small compared to the length scales on which the electric, \textbf{E}, and magnetic, \textbf{B}, fields vary, the cyclotron period must be short compared to the field variation time scales and the $\bf{E}\times\bf{B}$ drift velocity must be small compared to the particle thermal speed.

In spherical tokamaks (STs), such as the Mega Amp\`ere Spherical Tokamak (MAST) \cite{Meyer09}, the drift ordering is often not applicable, and it is then not possible to treat particle motion analytically. On the low field side of a ST plasma the poloidal component of the magnetic field can be comparable to the toroidal component, so that the drift orbit width of a trapped particle can be of the same order as the particle Larmor radius. Moreover, in ST plasmas with edge transport barriers (ETBs) the associated radial electric field can vary on length scales comparable to both the drift orbit width and the Larmor radius of thermal ions \cite{Meyer09}. 

It has been shown that both collisionless trapped particle guiding centre orbits and particle gyro-motion are significantly distorted by strongly-sheared radial electric fields \cite{Shaing92, Shaing98}. This effect is often referred to as \emph{orbit squeezing} \cite{Hazeltine89}, although in fact the widths of trapped particle orbits can be either reduced or increased, depending on the sign of the electric field gradient. The effects of orbit squeezing on neoclassical bulk ion transport have previously been studied in the limit of large aspect ratio \cite{Shaing92,Wang01} using this framework. However, the analytical theory of orbit squeezing assumes linearly sheared electric fields, and recent measurements of radial electric fields associated with ETBs in spherical tokamaks indicate field profiles that are not consistent with the assumption of constant shear \cite{Meyer09}. This highlights one limitation of the current theory and motivates the use of a numerical approach.

We note that non-uniformly sheared electric fields have previously been investigated using a full orbit approach \cite{Tao93}. However, this study investigated the impact of shear on particle $\bf{E}\times\bf{B}$ flow and only considered collisionless particle orbits. In the present work we investigate the impact on the motion and collisional transport of impurities of a strongly-sheared electric field in a MAST-like spherical tokamak equilibrium using a full orbit, test-particle simulation code. The fields are assumed to be axisymmetric and static; we therefore neglect turbulent transport effects.

\section{Model}

The transport of impurities in a spherical tokamak with a prescribed local radial electric field is simulated using the test particle code CUEBIT \cite{Hamilton03,McKay08}, which solves the Lorentz-Langevin equation
\begin{equation} \label{eq:LorentzLangevin}
    \mz \frac{\rmd\textbf{v}}{\rmd t} = Ze \left( \textbf{E} + \textbf{v} \times \textbf{B} \right) - \frac{\mz}{\tauzi}\left( \textbf{v} - \textbf{u} \right) + \mz \textbf{a}_r ,
\end{equation}
where $\mz$, $Ze$ and \textbf{v} are the impurity mass, charge and velocity, respectively. The term on the right-hand side of \eref{eq:LorentzLangevin} that is proportional to ${\bf v}-{\bf u}$ models the friction from the bulk ions. In the absence of an electric field it ensures that the impurities relax to a drifting Maxwellian distribution temperature $T_{\mathrm{i}}$ and flow \textbf{u} matching those of the bulk ions. Here, \textbf{u} is taken to be zero; we assume that the radial Lorentz force is sub-dominant to the pressure gradient term in the radial component of the bulk ion fluid momentum balance equation. This assumption is consistent with reported measurements of temperatures, densities and flows in the vicinity of ETBs in MAST \cite{Meyer09,Field09}: the pressure gradient is typically several tens of kPam$^{-1}$ whereas the contributions of poloidal and toroidal flows to the radial component of the Lorentz force are typically around one or two kPam$^{-1}$ at most. In taking \textbf{u} to be zero we also neglect any deviation of the bulk ion distribution from a stationary Maxwellian arising from neoclassical effects \cite{Helander02}. The quantity $\tauzi$ is the classical impurity-bulk ion collision time given by \cite{Helander02}
\begin{equation} \label{eq:collisionTime}
 \tauzi \equiv \frac{1}{\nuzi} = \frac{6\sqrt{2} \pi^{3/2} \epsilon^2_0}{\ln{\Lambda}} \frac{ \mz T_{\mathrm{i}}^{3/2}}{m^{1/2}_{\mathrm{i}} Z^2e^4 n_{\mathrm{i}}} ,
\end{equation}
which is a function of the local bulk ion temperature and density profiles, $T_i$ and $n_i$, respectively. Here, $\ln{\Lambda}$ is the Coulomb logarithm and $\epsilon_0$ is the permittivity of free space.

The final term on the right-hand side of \eref{eq:LorentzLangevin} models random accelerations of impurities due to Coulomb collisions with bulk ions. The vector $\textbf{a}_{\mathrm{r}}$ is a set of three random numbers, chosen independently for each particle at each time step from a Gaussian distribution of zero mean and variance
\begin{equation}
 \sigma^2 = \frac{2T_{\mathrm{i}}\nuzi}{\mz\Delta t} ,
\end{equation}
where $\Delta t$ is the time step that is used in the numerical simulation. Although the instantaneous collisions are taken to be isotropic, the cumulative effect of these over an impurity Larmor orbit naturally reflects the effects of gradients in the bulk ion temperature and density profiles (leading to, for example, thermal forces), which leads to anisotropy in the collisions. Here we have again assumed that the bulk ions have a Maxwellian distribution. 

Furthermore, in simulating test particles we assume that $Z^2 n_{\mathrm{z}} \ll n_{\mathrm{i}} \left( m_{\mathrm{e}} / m_{\mathrm{i}}\right)^{1/2}$, where $Z$ is the impurity ion charge number, $n_{\mathrm{z}}$ is the impurity ion number density, $n_{\mathrm{i}}$ is the bulk ion number density and $m_{\mathrm{e}}$ and $m_{\mathrm{i}}$ are the electron and bulk ion masses respectively \cite{Helander02}. This condition essentially ensures that collisions of bulk ions with impurity ions are sufficiently infrequent that they have a negligible effect on the bulk plasma neoclassical transport and automatically ensures that collisions between impurity ions can be neglected.

A MAST-like equilibrium is considered, which is modelled using an analytical Solov'ev-type solution of the Grad-Shafranov equation with a linear pressure profile and a potential toroidal magnetic field component. In cylindrical coordinates (R,$\phi$,Z), the poloidal flux, $\Psi$, is of the form
\begin{equation} \label{eq:solovev}
 \Psi = \Psi_0\left\{ \frac{\gamma}{8} \left[ \left( R^2 - R^2_0 \right)^2 - R^4_{\mathrm{b}}\right] + \frac{1-\gamma}{2}R^2 Z^2 \right\} ,
\end{equation}
with $R_0=0.964\mbox{ m}$, $R_{\mathrm{b}}=0.93\mbox{ m}$, $\gamma=0.8$ and $\Psi_0\simeq0.9\mbox{ Tm$^{-2}$}$ corresponding to a total plasma current of 1 MA. The magnetic field vector is calculated from \eref{eq:solovev} as $\textbf{B}$=$RB_\phi\nabla\phi$+$\nabla\Psi\times\nabla\phi$ where $RB_\phi$= 0.386 Tm. We assume $T_\mathrm{i}$ and $n_\mathrm{i}$ are flux functions and prescribe analytical forms,
\begin{eqnarray}
 T_\mathrm{i} & = & T_0\frac{\Psi}{\Psi_{\mathrm{m}}} + T_1, \\
 n_\mathrm{i} & = & n_0\frac{\Psi}{\Psi_{\mathrm{m}}} + n_1,
\end{eqnarray}
where $\Psi_{\mathrm{m}}$ the flux at the magnetic axis and $T_0=1$ keV, $T_1=0.5$ keV, $n_0=5\times 10^{19}$ m$^{-3}$ and $n_1=1\times 10^{19}$ m$^{-3}$. These profiles are broadly consistent with those measured in co-rotating MAST H-mode plasma discharges \cite{Akers04}. In such discharges, the effective ion charge is often measured to be close to unity throughout the core plasma, indicating no significant impurity accumulation.

\begin{figure}[htb]
 \centering
 \includegraphics[scale=0.75]{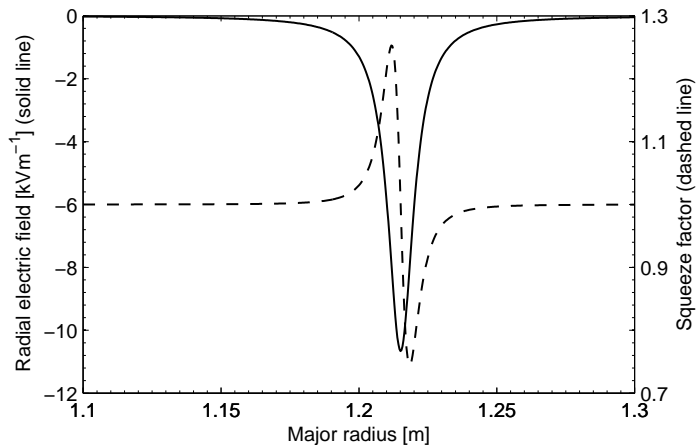}
  \caption{Form of the prescribed electric field (solid line) and corresponding squeeze factor for a C$^{6+}$ (dashed line) ion in the outer midplane of a MAST-like equilibrium. The full-width half-maximum of the radial electric field is chosen to be equal to the bulk ion gyroradius evaluated at the location of maximum electric field. The squeeze factor is calculated from $S=1+ Ze\Phi''(r)/m_{\mathrm{z}}\Omega^2_{\mathrm{p}}$.}
 \label{fig:eField}
\end{figure}

A local radially sheared electric field $\bf{E}=-\nabla\Phi$ near the edge is included, with an associated electrostatic potential $\Phi$ of the form
\begin{equation} \label{eq:potential}
\Phi\left(\Psi\right) = \Phi_0 \arctan\left(\frac{\Psi-\Psi_1}{\Delta\Psi}\right) ,
\end{equation}
where $\Phi_0$, $\Psi_1$ and $\Delta\Psi$ are the potential barrier height, location in $\Psi$ space and width in $\Psi$ space, respectively. Measurements reported by Meyer \etal \cite{Meyer09} of edge radial electric fields in MAST H-mode plasmas (see for example, figure 7 of \cite{Meyer09}) show that the corresponding potential profile is approximately of the form given by \eref{eq:potential} and that these constants have typical values of $\Phi_0$=60 V and $\Delta\Psi$=1.35 10$^{-3}$ Tm$^2$. We choose, for computational ease, to centre the electric field structure on $\Psi_1$=0.6$\Psi(R_0,0)$=-0.4 Tm$^2$. For our particular choice of equilibrium, these quantities correspond to a physical location and width in the outboard $Z$=0 plane of $R_1$=1.2 m and $\Delta{R}$=1.1 cm. The corresponding peak electric field strength, $E_0$, is $-10.5$ kVm$^{-1}$, which is close to measured values in MAST ETBs \cite{Meyer09}. We shall present results from simulations with varying values of $\Phi_0$ and $\Delta\Psi$ within the ranges $-4.75 \leq \Phi_0 \leq -120$ and $0.01\Psi(R_0,0) \leq \Delta\Psi \leq 0.08\Psi(R_0,0)$. The form of this prescribed electric field is shown in \fref{fig:eField}, which illustrates a limitation of using the orbit squeezing model: namely, which value of the squeeze factor to use when characterising the impact of the electric field on particle motion.

In addition to a radial electric field, the inward Ware pinch \cite{Ware70} is modelled approximately with a constant toroidal electric field $E_{\phi}=-0.3 \mbox{ Vm$^{-1}$}$. Including the $1/R$ dependence of this field has a small, $\Or(<10\%)$, effect on measured diffusion coefficients.

\section{Characterising transport}

\begin{figure}[htb]
 \centering
 \includegraphics[scale=0.75]{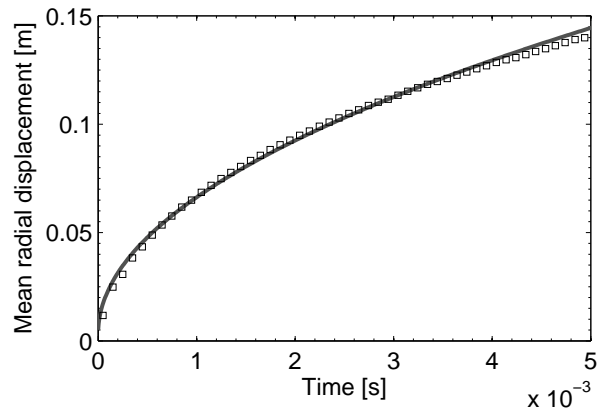}
  \caption{Mean minor radial coordinate versus time of C$^{6+}$ ions released at the magnetic axis in a simulation with zero electric field (squares). Overlaid is a fitted $t^{1/2}$ curve expected of diffusive transport (solid line).}
 \label{fig:meanDisplacement}
\end{figure}

In the absence of an electric field we expect neoclassical particle transport to be purely diffusive \cite{Helander02}. For tokamak plasmas with slowly-varying profiles, transport coefficients may be deduced empirically from moments of the spatial distribution of test particles \cite{Wong89,McClements09}. \Fref{fig:meanDisplacement}, in which the evolution of the mean minor radius of the impurity distribution is plotted, illustrates a typical CUEBIT simulation in which $10^4$ {C}$^{6+}$ ions are initially released from the magnetic axis in the absence of any electric field. Overlaid is a $t^{1/2}$ fit expected from diffusive transport \cite{Callaghan09}. This corresponds to a diffusion coefficient of ${D}\approx10\mbox{ m$^2$s$^{-1}$}$, which is close to the expected Pfirsch-Schl\"uter neoclassical diffusivity in the plasma core, $D \sim q^2 \rho_{\mathrm{L}}^2 /\tauzi$. Here $q$ is the safety factor and $\rho_{\mathrm{L}}$ is the Larmor radius.

In the case of narrow transport barriers, with plasma properties varying on length scales down to the Larmor radius, it is not possible to infer local transport coefficients using this method. In such cases a local effective diffusivity $\Deff$ may be deduced directly from the local radial particle flux $\Gamma_{\mathrm{z}}$ and density gradient $\partial n_{\mathrm{z}}/\partial\rho$:
\begin{equation} \label{eq:effectiveD}
 \Deff = - \frac{\Gamma_{\mathrm{z}}}{\partial n_{\mathrm{z}} / \partial \rho} .
\end{equation}
Here $n_{\mathrm{z}}$ is the flux surface-averaged minority ion density and $\rho=\rho(\Psi)$ is a radial coordinate, defined as the flux surface averaged minor radial coordinate
\begin{equation}
 \rho(\Psi) = \int_{\Psi_{\mathrm{m}}}^{\Psi} \frac{\int{Rdl/|\nabla\Psi'|}}{\int{Rdl}} d\Psi' ,
\end{equation}
with $l$ the arc length along a flux surface in the R--Z plane. $\rho(\Psi)$ is evaluated numerically as a function of $\Psi$ for the Solov'ev equilibrium described above. To quantify the impact of sheared electric fields on particle transport the value of $\Deff$ is normalised to its value measured in the absence of any electric field, $D$.

Clearly $\Deff$ is a simplified formulation of the particle diffusion coefficient since we assume that particle flux scales linearly with the minority ion density gradient. However, we wish to emphasise that $\Gamma_{\mathrm{Z}}$, which appears in \eref{eq:effectiveD}, is the full particle flux as computed by CUEBIT. This includes the effects of transport driven by bulk ion gradients, e.g. the thermal force arising from the bulk ion temperature gradient (see, for example, equation 5.9 of \cite{Helander02}). Such a force arises from the variation of the collision frequency across a Larmor orbit and, unlike to guiding centre and fluid calculations, appears naturally in full orbit particle simulations.

In order to elucidate the impact of radially sheared electric fields on particle transport we use \eref{eq:LorentzLangevin}. In order to proceed analytically we neglect both the random Langevin term and the bulk ion flow velocity, whose contribution to the momentum balance equation can, as noted previously, be assumed to be sub-dominant to the bulk ion pressure gradient term. The combined effect of the electric field and collisional drag terms can be seen by adopting Cartesian coordinates, with $\textbf{B}=B_z$ and $\textbf{E}=E_x$ both assumed to be constant, where we use the $x$-direction as a proxy for the radial direction in tokamak geometry. \Eref{eq:LorentzLangevin} can then be solved exactly, yielding
\begin{eqnarray}
 v_{\mathrm{x}} &=& v_{\perp} \exp{(-\nuzi t)} \sin\Omega t + \frac{E_{\mathrm{x}}}{B} \frac{\nuzi}{\Omega} \frac{1}{1+\nuzi^2/\Omega^2} , \\
 v_{\mathrm{y}} &=& v_{\perp} \exp{(-\nuzi t)} \cos\Omega t - \frac{E_{\mathrm{x}}}{B}\frac{1}{1+\nuzi^2/\Omega^2} .
\end{eqnarray}
In tokamaks we have that $\nuzi^2 \ll \Omega^2$. In this limit, we are left with the usual $\textbf{E}\times\textbf{B}$ drift in the $y$-direction and an additional ``drag force" drift in the $x$-direction, which arises from the inclusion of a collisional drag in the test particle equation of motion. This implies an inward-pinch in the presence of a negative radial electric field. Substituting for the impurity-ion collision frequency, given by \eref{eq:collisionTime}, we find that the drag force drift can be written as
\begin{equation} \label{eq:dragForceVelocity}
 v_{\mathrm{d}} \simeq {Z} \frac{E_{\mathrm{x}}}{B^2} \frac{n_{\mathrm{i}}}{T_{\mathrm{i}}^{3/2}} \frac{ m^{1/2}_{\mathrm{i}} e^3 \ln{\Lambda}}{6\sqrt{2} \pi^{3/2} \epsilon^2_0} .
\end{equation}
This is independent of test particle mass but proportional to test particle charge.

In order to relate the impact of a collisional drag force drift on particle transport we consider the steady-state one-dimensional transport equation
\begin{equation} \label{eq:denTransport}
\frac{\rmd}{\rmd x} \left\{ D\frac{\rmd n_{\mathrm{z}}}{\rmd x} + \frac{v n_{\mathrm{z}}}{1+(x/\Delta x)^2} \right\} = 0 ,
\end{equation}
where $v$ is the peak value of the drag drift, $v_{\mathrm{d}}$. Here we have assumed a form of the spatial variation of $E_{\mathrm{x}}$ that is consistent with the electrostatic potential given in \eref{eq:potential}, with $x$ the displacement from the peak electric field ($x \equiv \bar{\Psi}-\bar{\Psi}_1$) and $\Delta x$ a measure of the electric barrier width ($\Delta x\equiv \Delta\bar{\Psi}$). We neglect all spatial variations in $v_{\mathrm{d}}$ except for that occurring due to its dependence on $E_{\mathrm{x}}$. We also assume that the impurity ion Larmor radii are small compared to $\Delta x$ and that particle drifts in the $x$-direction are due solely to the drag effect. The above form of the transport equation ensures that the test particle flux, which we assume is composed of the usual diffusive and advective terms,
\begin{equation} \label{eq:particleFlux}
 \Gamma_{\mathrm{z}} = - D \frac{\rmd n_{\mathrm{z}}}{\rmd x} - v_{\mathrm{d}} n_{\mathrm{z}} ,
\end{equation}
is constant. Introducing a dimensionless spatial variable $\xi = x/\Delta x$ and a P\'eclet number $\mbox{Pe} = v\Delta x/D$, \eref{eq:denTransport} becomes
\begin{equation}
\frac{\rmd n_{\mathrm{z}}}{\rmd\xi} + \frac{\mbox{Pe} n_{\mathrm{z}}}{1+\xi^2} = -\gamma ,
\end{equation}
where $\gamma = \Gamma_{\mathrm{z}}\Delta x/D$. This has the exact solution
\begin{equation}
n_{\mathrm{z}}(\xi) = n_0 \exp{\left(-\mathrm{Pe}\tan^{-1}\xi\right)} \left[1 -\gamma\int_0^{\xi}\exp{\left(\mathrm{Pe}\tan^{-1}\eta\right)} \rmd\eta \right]
\end{equation}
where $n_0$ is the particle density at $x=0$, i.e. at the position of peak electric field and drag drift. For $\xi \ll 1$, or $x \ll \Delta x$, this reduces to
\begin{equation}
 n_{\mathrm{z}}(x) = n_0 \exp{\left(-\frac{\mathrm{Pe} x}{\Delta x}\right)}-\frac{\Gamma_{\mathrm{z}}}{v}\left[1-\exp{\left(-\frac{\mathrm{Pe} x}{\Delta x}\right)}\right].
\end{equation}
It follows that, at $x=0$, we have
\begin{equation}
 \Deff \equiv -{\Gamma_{\mathrm{z}}\over dn_{\mathrm{z}}/dx} = {D\over 1+vn_0/\Gamma_{\mathrm{z}}}.
\end{equation}
Identifying $v$ with the drag force drift velocity, and assuming that $n_{\mathrm{z}}/\Gamma_{\mathrm{z}}$ is essentially constant between simulations, we therefore find that the local effective particle diffusion coefficient scales with test particle charge as
\begin{equation} \label{eq:diffScaling}
 \Deff = \frac{D}{1 + \alpha Z} ,
\end{equation}
where 
\begin{equation}
\alpha = \frac{E_{\mathrm{x}}}{B^2} \frac{n_0n_{\mathrm{i}}}{T_{\mathrm{i}}^{3/2}} \frac{ m_{\mathrm{i}}^{1/2}e^3 \ln{\Lambda} }{6\sqrt{2}\pi^{3/2}\epsilon_0^2 \Gamma_{\mathrm{z}}}, 
\end{equation}
and is independent of both the test particle mass and charge. Despite the approximations used to derive \eref{eq:diffScaling}, we will show in the next section that it provides a fairly accurate description of impurity ion transport at an ETB in full toroidal geometry.

\section{Impurity transport simulations}

\subsection{Transport scaling with particle parameters}

We now present simulation results for a number of impurity species. In \tref{tab:collFrequencies} we list, for reference, the normalised collision frequency,
\begin{equation}
 \nu^* \equiv \frac{qR}{\tauzi v_{\mathrm{th}}} ,
\end{equation}
of each species at the position of the simulated transport barrier. Here $v_{\mathrm{th}}$ is the particle thermal velocity. The quantity $\nu^*$ measures the number of collisions a particle undergoes in one toroidal transit of the tokamak, with $\nu^* \leq 1$ indicating that a species is in the banana regime and $\nu^* > 1$ that a species is in the Pfirsch-Schl\"{u}ter regime.

\Table{\label{tab:collFrequencies}Normalised collision frequency for impurity ions simulated with computed effective diffusion coefficients, \eref{eq:effectiveD}, without, $D_0$, and with, $D_{\mathrm{eff}}$, a sheared radial electric field as given by \eref{eq:potential}. For those simulations indicated by $^*$ we have modified the location of the barrier, such that $\Psi_1=0.3\Psi(R_0,0)=-0.2$ Tm$^2$, and temperature and density profiles, such that $T_0=0.75$ keV, $T_1=0.05$ keV, $n_0=5\times 10^{19}$ m$^{-3}$ and $n_1=2\times 10^{19}$ m$^{-3}$.}
\br
 Impurity species & $\nu^*$ & $D_0$ & $D_{\mathrm{eff}}$   \\
\mr
  He$^{2+}$  &  0.032 & $29.7 \pm 1.4$ & $11.7 \pm 0.8$    \\
  C$^{6+}$   &  0.167 & $ 7.8 \pm 0.5$ & $ 0.8 \pm 0.2$    \\
  Ne$^{2+}$  &  0.014 & $43.7 \pm 0.8$ & $33.0 \pm 0.5$    \\
  Ne$^{4+}$  &  0.057 & $18.5 \pm 0.6$ & $10.0 \pm 0.3$    \\
  Ne$^{6+}$  &  0.129 & $10.7 \pm 0.7$ & $ 4.1 \pm 0.2$    \\
  Ne$^{8+}$  &  0.229 & $ 6.3 \pm 0.4$ & $ 2.2 \pm 0.2$    \\
  Ne$^{10+}$ &  0.358 & $ 4.5 \pm 0.3$ & $ 1.5 \pm 0.1$    \\
             &  4.737 & $ 2.8 \pm 0.3$ & $0.75 \pm 0.07^*$ \\
  Ar$^{10+}$ &  0.253 & $ 5.3 \pm 0.3$ & $ 1.6 \pm 0.1$    \\
             &  3.349 & $ 2.7 \pm 0.2$ & $0.49 \pm 0.05^*$ \\
  Mo$^{10+}$ &  0.167 & $ 6.4 \pm 0.8$ & $ 1.7 \pm 0.08$   \\
             &  2.208 & $ 3.4 \pm 0.2$ & $0.36 \pm 0.03^*$ \\
  W$^{10+}$  &  0.118 & $ 6.2 \pm 0.2$ & $ 2.2 \pm 0.07$   \\
             &  1.561 & $ 3.5 \pm 0.1$ & $0.29 \pm 0.02^*$ \\
  W$^{20+}$  &  0.473 & $ 1.7 \pm 0.1$ & $ < 0.2 $         \\
\br
\endTable

\begin{figure}[htb]
 \centering
 \includegraphics[scale=1.0]{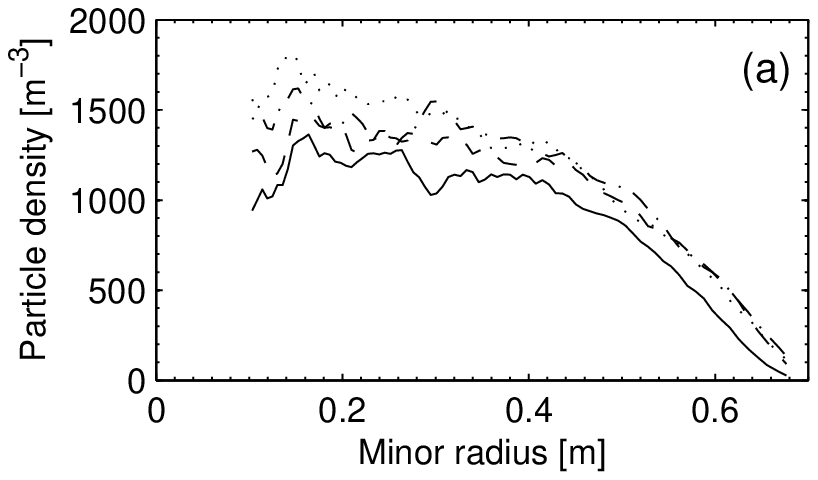}
 \includegraphics[scale=1.0]{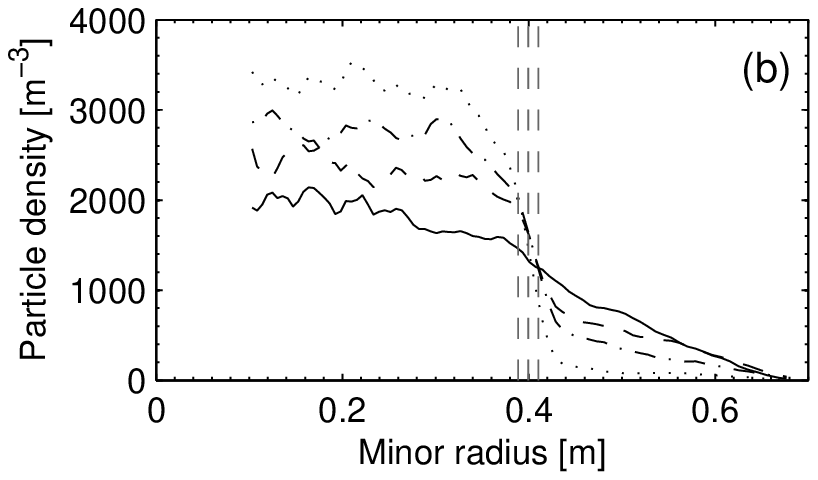}
 \caption{Density profiles of impurity ions against minor radius, both without (a) and with (b) a sheared radial electric field. Simulated ions are {H}e$^{2+}$ (\full), {C}$^{6+}$ (\broken), {N}e$^{10+}$ (\chain) and {W}$^{20+}$ (\dotted). Vertical lines in (b) indicate the peak strength and full-width half-maximum positions of the sheared radial electric field.}
 \label{fig:densityProfiles}
\end{figure}

\Fref{fig:densityProfiles} shows the results from simulations in which $10^4$ ions of various impurity species were released from the magnetic axis in equilibria with and without an inward-directed, radially sheared electric field. The electric field acts to confine {C}$^{6+}$ ions. The comparison of density profiles of {C}$^{6+}$ [see dashed lines, \fref{fig:densityProfiles}(a) and \fref{fig:densityProfiles}(b)] shows a step in the density profile at a minor radius of approximately 0.4 m. This indicates that the sheared electric field is acting as a barrier to the outward radial transport of the ions. The location of this step coincides with the peak of the radial electric field.

\begin{figure}[htb]
 \centering
 \includegraphics[scale=0.75]{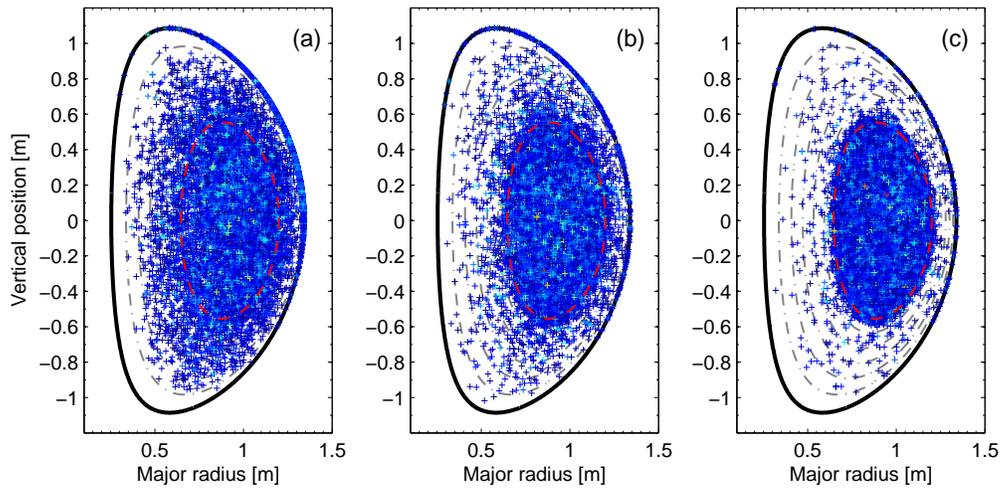}
 \caption{(Colour online) Poloidal distribution of (a) {H}e$^{2+}$ (b) {N}e$^{10+}$ and (c) {W}$^{20+}$ ions in a MAST-like equilibrium with a radially sheared electric field, the position of which is indicated by a dashed (red online) line.}
 \label{fig:polConfinement}
\end{figure}

The variation of impurity density with minor radial coordinate for three other impurity species, {H}e$^{2+}$, {N}e$^{10+}$ and {W}$^{20+}$, are shown on \fref{fig:densityProfiles}. The density profiles are similar in the absence of an electric field but differ substantially when it is present. The greatest confinement occurs for {W}$^{20+}$. All impurity species simulated have reached thermal equilibrium with the bulk ions particles before they begin interacting with the radial electric field. The poloidal distribution after 100 ms of simulated time are shown in \fref{fig:polConfinement} for {H}e$^{2+}$, {N}e$^{10+}$ and {W}$^{20+}$ ions, which again clearly illustrates the enhanced confinement of impurity ions by a sheared radial electric field, particularly in the case of {W}$^{20+}$.

\begin{figure}[htb]
 \centering
 \includegraphics[scale=0.75]{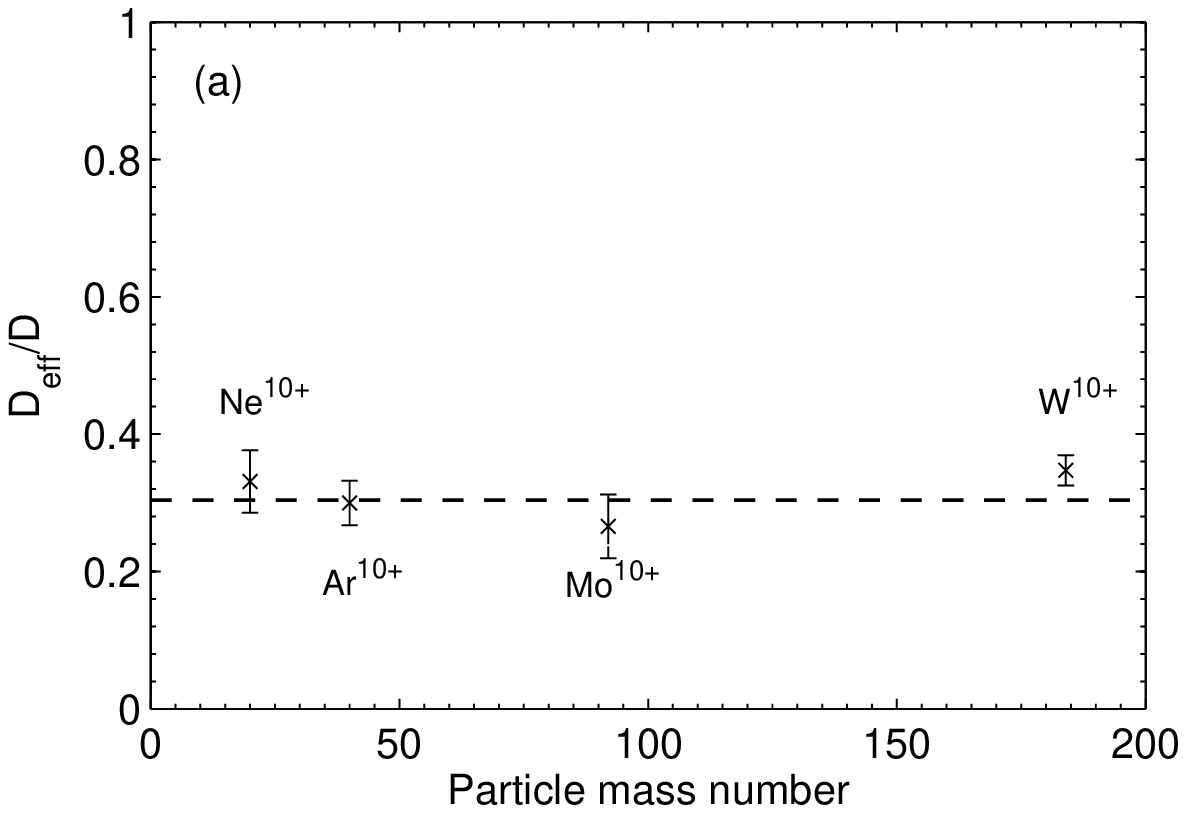}
 \includegraphics[scale=0.75]{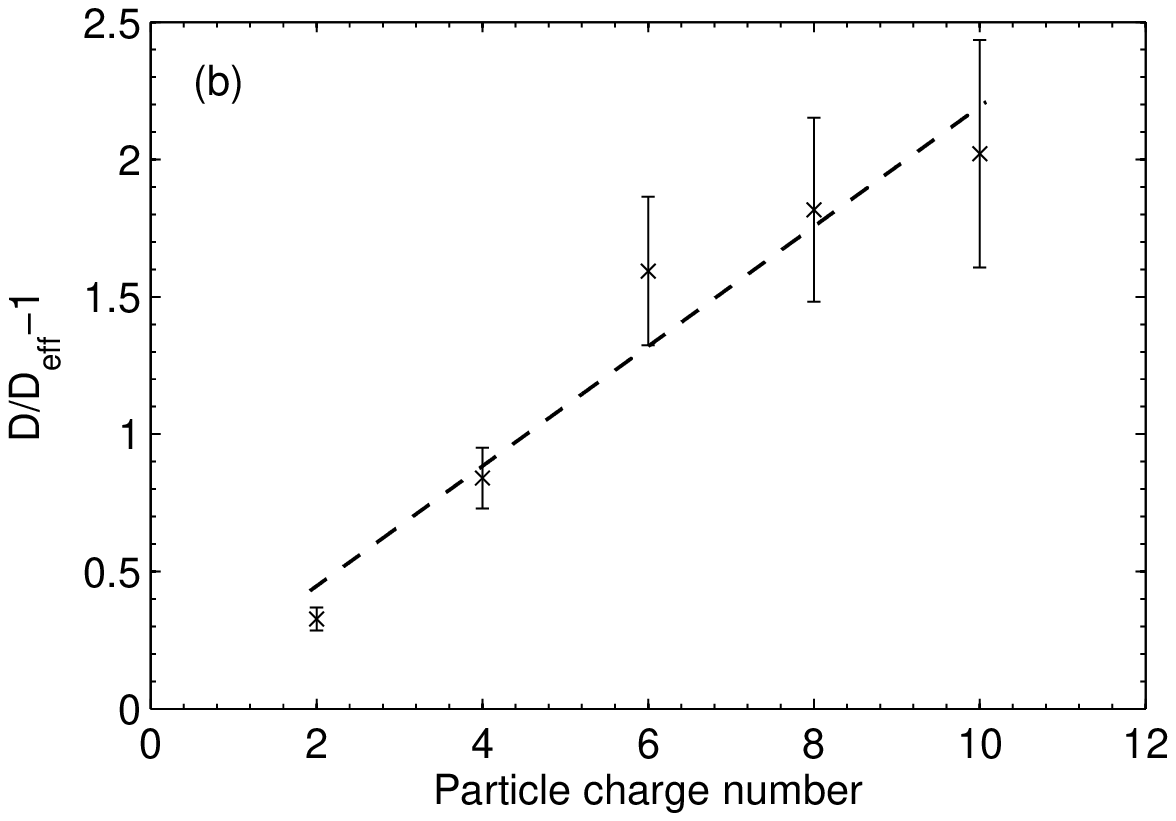}
 \caption{Scaling of particle diffusion coefficient with particles parameters. (a) scaling with test particle mass with constant charge number Z=10. (b) scaling with test particle charge for constant mass number of 20 (neon).}
 \label{fig:particleScaling}
\end{figure}

In order to compare the scaling of effective diffusivity with particle parameters, given by \eref{eq:diffScaling}, with simulations we consider results for a number of test particle species with varying mass and charge numbers. In each case the simulation is run until the ratio given by \eref{eq:effectiveD} at the peak of the electric field has reached a steady-state, except for noise fluctuations arising from the use of a finite number of test particles. However, the absence of a continuous particle source means that both the particle flux and the density gradient continue to slowly decline. In \fref{fig:particleScaling}a the normalised diffusion coefficient is plotted for particle mass numbers in the range 20 (neon) to 184 (tungsten) all with Z=10. We see that the diffusion coefficient is essentially independent of mass number. In \fref{fig:particleScaling}b the test particle mass number is held fixed at 20 but the charge state is varied from Z=2 to Z=10. \Eref{eq:diffScaling} predicts that the quantity $D/\Deff-1$ varies linearly with particle charge number. This is broadly consistent with the results shown in \fref{fig:particleScaling}b, although the uncertainties in $D/\Deff-1$ increase with $Z$. Furthermore, in deriving \eref{eq:diffScaling} contributions to the particle flux from bulk ion density and temperature gradients were neglected (cf. \eref{eq:particleFlux} and equation 5.9 of \cite{Helander02}). Including such terms would add to \eref{eq:diffScaling} a term that is non-linear in $Z$ in its denominator and which could explain the possible turn-over present for high $Z$ in \fref{fig:particleScaling}.

\subsection{Transport scaling with electric field strength and width}

\begin{figure}[htb]
 \label{fig:contours}
 \centering
 \includegraphics[scale=1.0]{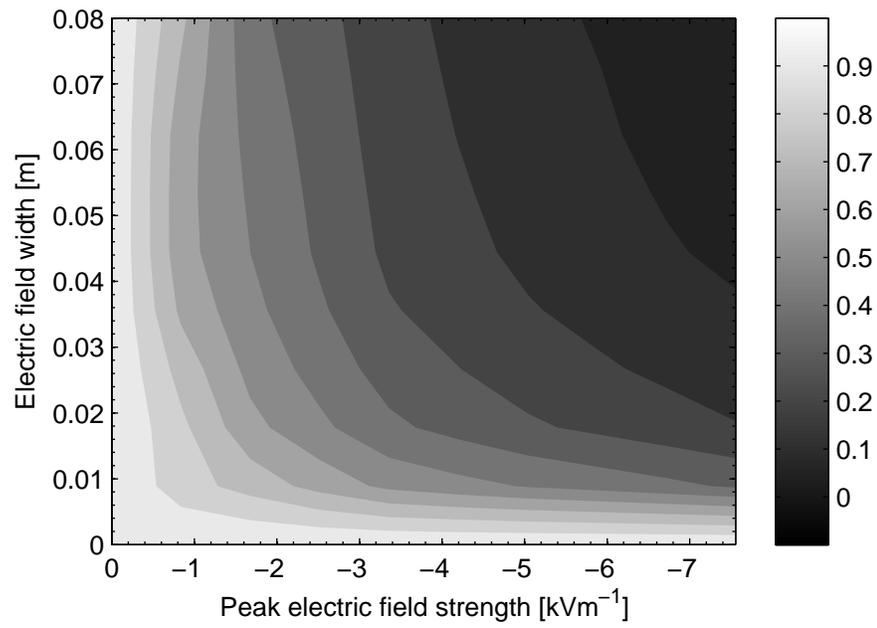}
 \caption{Contours of constant normalised effective diffusion coefficient for C$^{6+}$ ions with radial electric field height and width.}
 \label{fig:fieldScaling}
\end{figure}

\Fref{fig:fieldScaling} shows the variation with electric field width and height of the effective diffusion coefficient for C$^{6+}$ ions. The normalised effective diffusivity has a power law scaling with electric barrier width for constant barrier height. The coefficients of the fit themselves scale linearly with barrier height. Thus the scaling of diffusion coefficient with electric field width and strength may be expressed in the form
\begin{equation} \label{eq:empirical}
 \Deff = D_0 \left[ \frac{\Delta R}{\Delta R_0} \right]^{E/E_0} \mbox{ m$^2$s$^{-1}$} ,
\end{equation}
where $\Delta R$ is the FWHM of the electric field in the outer midplane and $E$ is the peak height of the radial electric field barrier. We see that the diffusion coefficient scales as the width of the electric field barrier to a power of the electric field strength, with the width of the barrier normalised to a fraction of the bulk ion Larmor radius. This scaling has been verified for $\Delta R$ and $E$ in the ranges $0.5\rho_{\mathrm{Li}} \leq \Delta R \leq 4\rho_{\mathrm{Li}}$ and $-0.65$ kVm$^{-1} \leq E \leq -7.65$ kVm$^{-1}$ for all impurity species, with $\rho_{\mathrm{Li}}$ the bulk ion Larmor radius. In the limit of no electric field, this scaling recovers the effective diffusivity as found in simulations without an electric field barrier. The scaling parameters for various species are listed in \tref{tab:escaling}.

Unlike \eref{eq:diffScaling}, the empirical expression for $\Deff$ given by \eref{eq:empirical} does in fact depend on the barrier width, implying limits in the validity of the approximations used to derive the former. We expect that for large $\Delta R / \rho_{\mathrm{Li}}$, the effective diffusion coefficient will no longer have a power law scaling but will tend asymptotically to a result similar to \eref{eq:diffScaling}.

We note, from \tref{tab:escaling}, that $ZE_0$ and $\Delta R_0$ are broadly constant. Therefore, the dependency of impurity transport on the shape of the electric field barrier depends approximately only on the charge state of the species, not on the species mass. Thus, we can approximate this scaling as
\begin{equation}
 \Deff \simeq D_0 \left( \frac{\Delta R}{0.2\rho_{\mathrm{Li}}} \right)^{ZE/57} .
\end{equation}

\Table{\label{tab:escaling}Scaling of computed effective diffusion coefficient, \eref{eq:effectiveD}, with electric field strength, $E$, and width, $\Delta R$, for several impurity species.}
\br
 Impurity species & $D_0$ (m$^2$s$^{-1}$) & $\Delta R_0$ & $E_0$ (kVm$^{-1}$)   \\
\mr
  C$^{6+}$   & $8.3 \pm 0.4$ & $\left(0.17 \pm 0.02 \right)\rho_{\mathrm{Li}}$ & $8.1 \pm 0.4$ \\
  Ne$^{10+}$ & $3.4 \pm 0.4$ & $\left(0.21 \pm 0.03 \right)\rho_{\mathrm{Li}}$ & $5.7 \pm 0.4$ \\
  Ar$^{15+}$ & $2.0 \pm 0.6$ & $\left(0.20 \pm 0.02 \right)\rho_{\mathrm{Li}}$ & $4.2 \pm 0.4$ \\
  Mo$^{20+}$ & $1.6 \pm 0.8$ & $\left(0.22 \pm 0.03 \right)\rho_{\mathrm{Li}}$ & $2.9 \pm 0.6$ \\
\br
\endTable

\section{Discussion}

We have simulated realistic radial electric field profiles for ETBs in a spherical tokamak equilibrium and investigated the impact on collisional impurity transport of the strength and width of the electric field profile for a variety of impurity species. We have demonstrated that full orbit simulations provide a valuable insight into the transport of particles in such scenarios. It has been shown that a strongly-sheared radial electric field can significantly increase the confinement of test particle ions. This effect has been explained in terms of a drift arising from the drag between impurity and bulk ions, with the direction of the drift given by the radial electric field and its strength determined by the impurity charge. Clearly, the negative implication of this study is the enhanced confinement of highly ionized impurity species in the vicinity of strong, negative radial electric fields.

In our derivation of the drag force drift given by \eref{eq:dragForceVelocity} we neglected both spatial variations in the radial electric field and toroidal geometry. However, the scaling of effective diffusivity with $Z$ inferred from using this drift velocity in a simple slab model of the transport barrier, \eref{eq:diffScaling}, is in agreement with our test-particle simulations in which no approximations are made with regard to finite Larmor radius or toroidal effects. This suggests that test-particle transport in the immediate vicinity of the barrier is not strongly-dependent on the effects of toroidal geometry or the collisionality regime (i.e. the value of $\nu^*$). This further indicates that finite Larmor radius effects are relatively unimportant. One would expect electric field shear on length scales approaching the particle Larmor radius to result in a modified drag force drift velocity, analogous to the $\bf{E}\times\bf{B}$ drift with finite Larmor radius correction discussed by Tao \etal \cite{Tao93}. However, this requires further study.

The location of the electric field in the simulations, which resembles more an internal transport barrier than an ETB, was chosen in order to lower computation time by reducing the distance from the plasma core to the location of the electric field, thereby reducing the time before particles begin interacting with the electric field. This also has the benefit of improving particle statistics, since the particles are less dispersed radially and the number of particles at the barrier is consequently higher. Simulations in which the electric barrier is located closer to the plasma edge, listed in \tref{tab:collFrequencies}, demonstrate that moving the barrier closer to the edge has little qualitative impact on the results obtained. Furthermore, the variation of collisionality of impurity species, with normalised collisionality ranging from $0.014$ (banana regime) to $4.74$ (Pfirsch-Schl\"uter regime), appears to be unimportant, since the measured normalised diffusivity remains independent of mass and scales as expected from the drag force drift argument with particle charge. 

Whilst the effect of bulk plasma rotation was neglected in the present study, one would expect transport barriers to be associated with flows of the bulk ions (e.g. zonal flows). For impurity ions in particular inertial effects will become important in rotating plasmas. As recently shown \cite{McClements09}, strong toroidal rotation can cause heavy, incompletely-ionised impurities such as W$^{20+}$ to undergo very rapid collisional transport (due to the combined effect of trapping in a centrifugal potential well and a modification to the effective magnetic field arising from the Coriolis force). One would therefore expect a competition between this enhanced transport due to rotation and the improved confinement resulting from sheared radial electric fields.

We note that the effect of improved confinement due to a sheared radial electric field is more important for spherical tokamak plasmas than for conventional tokamaks, primarily due to the strong $B$ dependence in \eref{eq:dragForceVelocity}. For example, in the MAST-like equilibrium used throughout, we have that the Ware pinch velocity is of the order $v_{\mathrm{Ware}} \simeq qE_{\phi}/\epsilon B \sim 5.6 \mbox{ ms}^{-1}$ and the drag force drift velocity is of the order $v_{\mathrm{d}} \sim 8.2$ ms$^{-1}$ for C$^{6+}$ (where we have used $B=0.4$ T, $E_x=-10$ kVm$^{-1}$, $E_{\phi}=-0.3$ Vm$^{-1}$, $n_i=3\times 10^{19}$ m$^{-3}$, $T_i=500$ eV, $q=3$ and $\epsilon =0.4$). Typical values for a large conventional tokamak such as JET, on the other hand, are $v_{\mathrm{Ware}}\sim 0.4$ ms$^{-1}$ and $v_{\mathrm{d}}\sim 0.02$ ms$^{-1}$ (assuming $B = 3\,$T, $E_x = 20\,$kVm$^{-1}$, $E_{\phi}=-0.04$ Vm$^{-1}$, $n_i=3\times 10^{19}$ m$^{-3}$, $T_i=5$ keV, $q=3$ and $\epsilon =0.1$ \cite{Tala07}). Thus, the drag force drift is much less significant than the Ware pinch under typical conventional tokamak conditions.

An area of possible future investigation is the impact of fluctuating radial electric fields on collisional particle transport, with the fluctuations resulting from, for example, geodesic acoustic mode (GAM) oscillations \cite{Kramer09}. Furthermore, in recent fluid simulations of the COMPASS-D tokamak, it has been observed that the H-mode radial electric field close to the plasma edge changes from being negative between edge localised modes (ELMs) to positive during an ELM \cite{Thyagaraja10}. In the context of the present work one would expect the direction of the drag force drift to switch between inward and outward directed, which could have important implications for impurity accumulation and the transport of particles at transport barriers. The flexibility of the full orbit approach means that turbulent electric or magnetic fields could be readily incorporated into numerical studies of impurity transport in ELMy H-mode plasmas.

\ack
This work was funded by the RCUK Energy Programme under grant EP/G003955 and a Science and Innovation award, and by the European Communities under the contract of Association between EURATOM and CCFE. The views and opinions expressed herein do not necessarily reflect those of the European Commission.

%--- Figures:

\end{document}